\documentclass[aps,prl,twocolumn,showpacs]{revtex4}
\usepackage{graphicx}
\begin{document}

\title{Quantum crystal phase of electron liquid}

\author{A. Kashuba}

\affiliation{L. D. Landau Institute for Theoretical Physics, Russian Academy of Sciences, 2 Kosygin str., 119334 Moscow}

\date{\today}

\begin{abstract}
A new bosonic, excitonic method for interacting electrons is developed. For two-dimensional electron liquid, it reveals a noncompensated quantum crystal phase ranging from the high density, or vanishing Coulomb interaction, limit: $r_s=0$, to a lower density $r_s\sim 10$, where a condensation of exciton pairs and the loss of the Fermi step is possible. The correlation energy and the static susceptibility, calculated using the method, are in agreement with the independent studies. A proof of the instability of the Fermi surface in the limit $r_s\rightarrow 0$ is given. The concept of quantum crystal is capable to explain the metal-insulator transition in Si field-effect heterostructures.
\end{abstract}


\maketitle

Electron liquid is a system of non-relativistic electrons with Coulomb interaction \cite{NP}.  Its quantum ground state depends on the density of electrons $n$ only, or on the normalized inter-electron distance $r_s$, with $r_s=e^2m/\sqrt{\pi n}$ in two dimensions (2D). Landau theory of Fermi liquid \cite{landau56,NP} and quantum Monte Carlo (QMC) method \cite{QMC,QMC1} predict that paramagnetic, uniform liquid, with the same type of excitations as in the free Fermi gas, is stable in 2D from $r_s=0$ to up to $r_s\sim 30$, whereas the experiment in 2D Si field-effect heterostructures \cite{krav94} finds a metal-insulator transition (MIT) at $r_s\sim 9$. Landau theory describes the Fermi liquid in terms of the two-particle scattering vertex $\Gamma$ \cite{AGD}. The translational, Galilean and gauge symmetries of the Fermi liquid impose constraints on the vertex $\Gamma$ \cite{AGD}, however, they alone are insufficient to determine it. While $\Gamma$ is unknown for arbitrary $r_s$ the conditions for the stability of the Fermi liquid is expressed in terms of $\Gamma$ \cite{pomer}. The QMC method \cite{QMC,QMC1} applied to Fermi liquid or Wigner crystal is limited by the use of variational wave function and, therefore, studies the phase predetermined by the wave function guess. The QMC method was incapable to reveal the instabilities of the Fermi liquid.

In this paper I propose a new method for uniform electron liquid (eliquid) where degrees of freedom are bosons representing excitons, the electron-hole pairs. The calculated correlation energy of 2D electron liquid as a function of $r_s$ agrees with the QMC results \cite{QMC1}. The method reproduces the Coulomb screening of the static potential \cite{NP}. New features found are two instabilities. First, the susceptibility on the Kohn momentum, twice the Fermi momentum $p_F$, diverges at $r_s\sim 2$ and this instability is independently proven to develop for any $r_s>0$. Second, the method reveals a Bose condensation of exciton pairs around $r_s\sim 10$, accompanied by the loss of the Fermi step discontinuity. Qualitative interpretation of these findings is that charge density waves develops opening gaps on the Fermi circle. The ground state of the 2D eliquid is the quantum crystal \cite{AL69} with, perhaps, triangular lattice and with the number of electrons that does not match the number of the lattice cites. Noncompensated electrons form two smaller, closed, hole-like Fermi circles. MIT is a point where the Fermi circles disappear and the quantum crystal transforms into the compensated Wigner crystal.

Only 2D spin polarized eliquid is considered here. The ground state of the corresponding Fermi gas is: $|G\rangle = \prod_{\mathbf{s}}\psi^+(\mathbf{s})\ |0\rangle $. Everywhere, $\mathbf{s}$ and $\mathbf{p}$ stands for a momentum inside and outside the Fermi circle. Accordingly, the Hamiltonian is the sum of i) the Hartree-Fock (HF) term:
\begin{equation}\label{HamiltonianHF}
\hat{H}_0=E_{HF}+\sum_{\mathbf{p}}\epsilon_{HF}(\mathbf{p})\psi^+_\mathbf{p}\psi_\mathbf{p}- \sum_{\mathbf{s}}\epsilon_{HF}(\mathbf{s})\psi_\mathbf{s}\psi^+_\mathbf{s}
\end{equation}
where $E_{HF}$ is the total HF energy, $\epsilon_{HF}(\mathbf{p})$ is the HF dispersion \cite{NP}, and the Coulomb terms: ii) creating or destroying two excitons:
\begin{equation}\label{Hamiltonian02}
\hat{H}^{02}=\!\!\!\!\!\sum_{\mathbf{s}_1<\mathbf{s}_2,\mathbf{p}_1<\mathbf{p}_2}\!\!\!\!\! U(\mathbf{p}_1\mathbf{p}_2; \mathbf{s}_2\mathbf{s}_1)  \left(\psi_{\mathbf{p}_1}^+ \psi_{\mathbf{p}_2}^+ \psi_{\mathbf{s}_2} \psi_{\mathbf{s}_1} + h.c. \right)
\end{equation}
where $U(\mathbf{p}_1\mathbf{p}_2; \mathbf{s}_2\mathbf{s}_1)=$  $\left(U(\mathbf{p}_1-\mathbf{s}_1)- U(\mathbf{p}_1-\mathbf{s}_2)\right)$ $\delta(\mathbf{p}_1+ \mathbf{p}_2-\mathbf{s}_1-\mathbf{s}_2)$, $U(\mathbf{q})= 2\pi e^2/|\mathbf{q}|\mathcal{A}$ is the Coulomb potential, $\mathcal{A}$ is the area filled by eliquid, iii) scattering quasiparticles while preserving the number of excitons:
\begin{eqnarray}\label{Hamiltonian22}
\hat{H}^{22}=\!\!\!\!\sum_{\mathbf{p}_1<\mathbf{p}_2,\mathbf{p'}_1<\mathbf{p'}_2}\!\!\!\! U(\mathbf{p}_1\mathbf{p}_2; \mathbf{p'}_2\mathbf{p'}_1) \psi^+_{\mathbf{p}_1} \psi^+_{\mathbf{p}_2} \psi_{\mathbf{p'}_2} \psi_{\mathbf{p'}_1}+ \nonumber\\ + \sum_{\mathbf{s}_1<\mathbf{s}_2,\mathbf{s'}_1<\mathbf{s'}_2} U(\mathbf{s}_1\mathbf{s}_2; \mathbf{s'}_2\mathbf{s'}_1) \psi_{\mathbf{s}_1} \psi_{\mathbf{s}_2} \psi^+_{\mathbf{s'}_2} \psi^+_{\mathbf{s'}_1}- \nonumber\\ - \sum_{\mathbf{p}_1\mathbf{p}_2\mathbf{s}_1\mathbf{s}_2} U(\mathbf{p}_1\mathbf{s}_2; \mathbf{s}_1\mathbf{p}_2) \psi^+_{\mathbf{p}_1} \psi_{\mathbf{s}_1} \psi^+_{\mathbf{s}_2} \psi_{\mathbf{p}_2}
\end{eqnarray}
and iv) creating or destroying one exciton:
\begin{eqnarray}\label{Hamiltonian12}
\hat{H}^{12}=\sum_{\mathbf{p}\mathbf{s}_3,\mathbf{s}_1<\mathbf{s}_2} U(\mathbf{p}\mathbf{s}_3; \mathbf{s}_2\mathbf{s}_1) \left( \psi^+_\mathbf{p} \psi_{\mathbf{s}_2} \psi_{\mathbf{s}_1} \psi^+_{\mathbf{s}_3} +h.c.\right)+ \nonumber\\
+\sum_{\mathbf{s}\mathbf{p}_3,\mathbf{p}_1<\mathbf{p}_2}U(\mathbf{p}_3\mathbf{s}; \mathbf{p}_2\mathbf{p}_1) \left( \psi^+_{\mathbf{p}_3} \psi^+_\mathbf{s} \psi_{\mathbf{p}_2} \psi_{\mathbf{p}_1} +h.c.\right)
\end{eqnarray}
All quantum states that emerge during the temporal evolution of eliquid with the Hamiltonian (\ref{HamiltonianHF}-\ref{Hamiltonian12}) are neutral and translationally invariant, i.e. with equal number of quasiparticles and quasiholes and with zero total excitation momentum. Among these excited states there is a set of irreducible states with the property that quasiparticles could not be partitioned into two neutral subsets with both having the zero total momentum. Let $(n,\mathbf{p}_1...\mathbf{p}_n, \mathbf{s}_1...\mathbf{s}_n)$ or $(n,i)$, with zero total momentum, be an index of the irreducible $n$-exciton state and
\begin{equation}\label{CreationOperator}
A^+_{ni}=\prod_{j=1}^n \psi_{\mathbf{p}_j}^+ \prod_{j=1}^n \psi_{\mathbf{s}_j}
\end{equation}
be the operator creating this state. They all commute:
\begin{equation}\label{Commutator}
[A^+_{ni}, A^+_{mj}]_- =0
\end{equation}
A state of eliquid is partitioned into the irreducible states very much like a number is a product of the prime numbers. In general this partition is not unique. For example, a four-exciton state with two excitons having momentum $\mathbf{q}$ and the other two excitons having momentum $-\mathbf{q}$ can be reduced into the irreducible two-exciton states in two different ways as shown in Fig.1A.

Consider a bosonic state specified by the index $(n,i)$ and the corresponding Bose operator $a^+_{ni}$. We associate with the action of operators $A$ in the electron Fock space the action of operators $A$ in the boson Fock space:
\begin{equation}
|exc\rangle =  A^+_{ni} \ |G\rangle  \ \leftrightarrow \ |0..1_{ni}..0\rangle = A^+_{ni} \ |0\rangle
\end{equation}
The boson Fock space is larger than the electron Fock space. First, the occupation number of any boson state can not exceed one. Second, a double excited state $|0..1_{ni}..1_{mj}..0\rangle = A^+_{ni}A^+_{mj} \ |G\rangle$, which is correctly defined because of Eq.(\ref{Commutator}), vanishes if at least one momentum enters both sets $(ni)$ and $(mj)$. Let $\Omega_\mathbf{p}$ be a set of the irreducible states $(n,i)$ that include the momentum $\mathbf{p}$. Then, the operator:
\begin{equation}\label{SchwingerRepresentation}
A^+_{ni}=a^+_{\mathbf{p}_1..\mathbf{p}_n,\mathbf{s}_1..\mathbf{s}_n}\prod_{j=1}^n b_{\mathbf{p}_j} b_{\mathbf{s}_j}
\end{equation}
conforms to the electron Fock space limits if the number of slave bosons $b_{\mathbf{p,s}}$ equals to one in the ground state $|G\rangle$:
\begin{equation}\label{SchwingerConstraint}
b^+_{\mathbf{p}}b_{\mathbf{p}}+\sum_{(n,i)\in \Omega_{\mathbf{p}}} a^+_{ni}a_{ni}=1
\end{equation}
This representation is similar to the Schwinger boson representation for a spin. The analog of the Holstein-Primakoff representation of a spin in our case reads:
\begin{equation}\label{HolsteinRepresentation}
A^+_{ni}=a^+_{ni}\prod_{j} \sqrt{1-\sum_{ml\in \Omega_{\mathbf{p}_j}}a^+_{ml}a_{ml}}  \sqrt{1-\sum_{ml\in \Omega_{\mathbf{s}_j}} a^+_{ml}a_{ml}}
\end{equation}
To avoid the double counting of the states in Fig.1A we exclude them in the $a$-boson space while a new boson: $c^+_{4m}$, where $(4m)$ is $(2i_1)(2i_2)$ or $(2j_1)(2j_2)$ identical because the momenta in these two sets are the same, will represent such a state. Let $\Omega_{\mathbf{q}}$ be a set of two-exciton states $(2i)$, with the momentum $\mathbf{q}$ in one of the two particle-hole channels $||$ or $\times$, and let $f_{\mathbf{q}}$ be the corresponding slave boson. Then, the operator
\begin{equation}\label{SchwingerRepresentationFull}
A^+_{2i}=a^+_{\mathbf{p}_1\mathbf{p}_2,\mathbf{s}_1\mathbf{s}_2} b_{\mathbf{p}_1} b_{\mathbf{s}_1} b_{\mathbf{p}_2} b_{\mathbf{s}_2} f_{\mathbf{q}_{||}} f_{\mathbf{q}_{\times}}
\end{equation}
and the constraint: $f^+_{\mathbf{q}}f_{\mathbf{q}}+\sum_{(2i)\in \Omega_{\mathbf{q}}} a^+_{2i}a_{2i}=1$, exclude the double-counting states. This constraint can be resolved to give the Holstein-Primakoff representation. Namely, it is Eq.(\ref{HolsteinRepresentation}) multiplied by two squareroots: $\prod_{||,\times}\sqrt{1-\sum_{(2m)\in \Omega_{\mathbf{q}}} a^+_{2m}a_{2m}}$, corresponding to the two double-counting channels. Representation of the higher many-exciton states may be more involved then the two-exciton state Eq.(\ref{SchwingerRepresentationFull}).

\begin{figure}\label{FigTwoWay}
\includegraphics[scale=0.65]{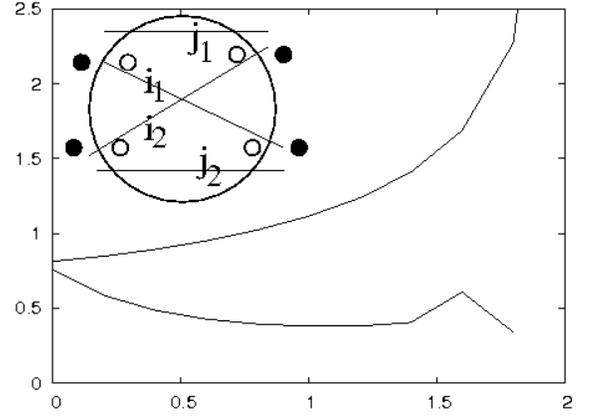}
\caption{A. Insert. The four-exciton state that can be reduced in the two different ways: $(2i_1),(2i_2)$ or $(2j_1),(2j_2)$. \\
B. Static susceptibility in units of $m/2\pi$ vs $r_s$, for $N=37$ and two wavevectors: the minimum on the lattice $Q\sim 0.3p_F$ and the nearest to the Kohn momentum: $Q\sim 1.86p_F$.}
\end{figure}

Our aim is to map the Hamiltonian (\ref{HamiltonianHF}-\ref{Hamiltonian12}) into the boson space. We do this by expanding the Hamiltonian order by order in number of excitons. This expansion is similar to $1/S$ expansion in the theory of antiferromagnet. It has been noted that $1/S$ expansion could be convergent \cite{Yang}, with the higher order terms contributing lesser corrections to the lower order terms, and this is due, perhaps, to the compactness of the spin manifold. Usual in the theory of Fermi liquid is the expansion in powers of $r_s$ \cite{NP}, and this, perhaps, is divergent, with long series and exact pattern of divergence required in order to make a useful resummation, like Borel in the theory of critical indices of the second order phase transition.

The boson Hamiltonian is expanded in terms ordered by the exciton number. First is the two-exciton term:
\begin{equation}
H_2=\!\!\!\! \sum_{\mathbf{p}_1<\mathbf{p}_2,\mathbf{s}_1<\mathbf{s}_2} U(\mathbf{p}_1,\mathbf{p}_2; \mathbf{s}_2\mathbf{s}_1) \left( a^+_{\mathbf{p}_1\mathbf{p}_2\mathbf{s}_2\mathbf{s}_1}+ a_{\mathbf{p}_1\mathbf{p}_2\mathbf{s}_2\mathbf{s}_1} \right)
\end{equation}
and next is the four-exciton term:
\begin{equation}
H_4=\sum_{ij}a^+_{2i}H^{22}_{ij}a_{2j}
\end{equation}
Scattering takes place in 6 different channels: particle-particle (first line of Eq.(\ref{Hamiltonian22})), hole-hole  (second line of Eq.(\ref{Hamiltonian22})), and four different particle-hole channels (third line of Eq.(\ref{Hamiltonian22})). $H^{22}$ matrix includes the fermionic sign of the momenta ordering before and after scattering.

Minimum of the Hamiltonian $H_2+H_4$ with respect to $a$ gives the correlation energy of eliquid in the lowest order. Quantization of momentum on the triangular lattice is used. For number of electrons $N=31,37,55,61$ the ground state possesses $D_{6h}$ symmetry which reduces the dimension of the Fock space by twelve. The momentum $\mathbf{p}$ is restricted by a maximum value $Rp_F$, with typically $R=2 .. 5$. The dependence of $E_c$ on $R$ is smooth and fitted by the law: $E_c(N,R)= E_{c}(N)- \kappa/R^{\beta}$, with $\beta= 5..8$ for $r_s=0..5$. $E_c(N)$ as a function of $N$ shows wide 'mesoscopic' variations, related to the electrostatic energy of $N$ electrons put in the box with the size determined by the lattice spacing. $E_c(N)$ variations are correlated with the variations of the particle-hole gap on the Fermi circle $\Delta(N)$, with $\Delta(\infty)=0$. In the limit $r_s\ll 1$ we find $E_c(N)= E_{c0}(N)-\alpha \Delta(N)$, for the quasiparticle dispersion: $\epsilon=\sqrt{\xi^2+\Delta^2}$. Therefore, $\Delta(N)$ is also calculated and the best $\alpha$ is found to lessen the $E_c(N)$ variations. Then, $E_{c0}(N)$ is fitted by the law reflecting the Coulomb singularity: $E_{c0}(N)=E_{c}+ \gamma/\sqrt{N}$. $E_c$ is shown in Table I and it is within $10\%$ of the QMC result \cite{QMC1}. The next approximation to the bosonic Hamiltonian includes: $H_5$ term that scatters two excitons into three-exciton states and $H_6$ term that scatters between three-exciton states as well as accounts for nonlinear three two-excitons processes. The occupation numbers $\langle \psi^+_\mathbf{p}\psi_\mathbf{p} \rangle= \sum_{i\in\Omega_\mathbf{p}} \langle a^+_{2i}a_{2i} \rangle/2$ and the corresponding Fermi step discontinuity have been found, see Table I. The Fermi step discontinuity vanishes at $r_s$ close to the point of the two-exciton Bose condensation.

\begin{table}
\caption{Correlation energy per electron in atomic units and the Fermi step discontinuity}
\begin{ruledtabular}
\begin{tabular}{cccccc}
$r_s$  &  1 & 2  &  3 & 4 & 5 \\
$-E_c$ & 0.0235 & 0.0194 & 0.0167 & 0.0149 & 0.0135 \\
$\Delta n(p_F)$ & 0.95 & 0.88 & 0.80 & 0.72 & 0.62 \\
\end{tabular}
\end{ruledtabular}
\end{table}

\begin{figure}\label{FigInstability}
\includegraphics[scale=0.55]{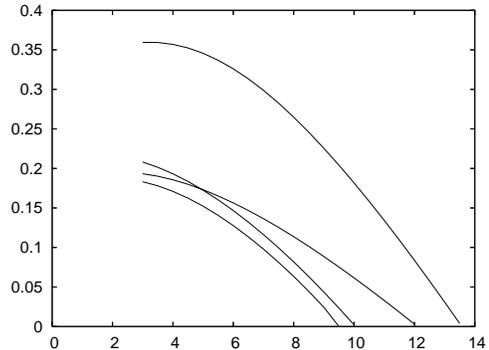}
\caption{The lowest energy in atomic units of two excitons as a function of $r_s$ for $N=61,55,85,37$ in the order of increasing $r_s$ of the condensation point}
\end{figure}

The lowest eigenvalue of the matrix $H^{22}_{ij}$ is shown in Fig.2. A bound state of two excitons with negative energy occurs at $r_s> 9..13$ with 'mesoscopic' variations depending on $N$ and excitons condense into this bound state. The configuration of two excitons, that contributes the most to the condensate wave function, has two holes with $\mathbf{s}_1=-\mathbf{s}_2$ and two particles with $\mathbf{p}_1=-\mathbf{p}_2$.

The response of eliquid to an external static potential $2V_\mathbf{q} \cos(\mathbf{q}\mathbf{r})$ can be also found using the boson representation. The potential breaks down the translational symmetry and, accordingly, an additional $\mathbf{q}$-boson space with one- and two-exciton operators $a_{1i\mathbf{q}}$ and $a_{2i\mathbf{q}}$ is introduced. For small $V_\mathbf{q}$, $\mathbf{q}$-states are slightly excited and the nonlinearity is negligible. To exclude the double counting of the states $a^+_{1i\mathbf{q}}a^+_{2j} |0\rangle $, with two excitons of the state $2j$ having momenta $\pm\mathbf{q}$, the operators $a^+_{2j}$ are modified as: $a^+_{2j}\sqrt{1-\sum_i a^+_{1i\mathbf{q}} a_{1i\mathbf{q}} }$. The $\mathbf{q}$-Hamiltonian for the eliquid response in the four-exciton order consists of the Coulomb and the potential terms:
\begin{eqnarray}
H^4_{\mathbf{q}}=\sum_{ij}a^+_{1i\mathbf{q}} H^{11}_{ij\mathbf{q}} a_{1j\mathbf{q}}+ \sum_{ij} a^+_{2i\mathbf{q}}H^{22}_{ij\mathbf{q}} a_{2j\mathbf{q}} +\nonumber\\
+\sum_{ij} \left( a^+_{2i\mathbf{q}} H^{21}_{ij\mathbf{q}} a_{1j\mathbf{q}}+h.c.\right)+ V_\mathbf{q}\sum_i \left(a^+_{1i\mathbf{q}}+ a_{1i\mathbf{q}} \right)+ \nonumber\\
+ V_\mathbf{q} \sum_{ij} \left(a^+_{2j} a_{1i\mathbf{q}}+ a^+_{2i\mathbf{q}} \delta^{22}_{ij\mathbf{q}} a_{2j} + h.c. \right)
\end{eqnarray}
where $\delta_\mathbf{q}$ is the momentum shift by $\mathbf{q}$. Integrating out the variables $a_{2i}$ from the Hamiltonian $H_2+H_4+H^4_\mathbf{q}$ we find an effective $\mathbf{q}$-Hamiltonian in the $\mathbf{q}$-space. Its minimum with respect to $a_{1i\mathbf{q}}$ and $a_{2i\mathbf{q}}$ gives the energy gain: $E_\mathbf{q}=-\chi(\mathbf{q}) V_\mathbf{q}^2$. The susceptibility $\chi(\mathbf{q})$ as a function of $r_s$ is shown in Fig.1B. In the long range limit $q\ll p_F$ it, the lower curve, is suppressed by the Coulomb screening: $\chi(q)\sim q/r_s p_F$, whereas the Kohn susceptibility $\chi(2p_F)$, the upper curve, diverges at $r_s\sim 2$. This divergence is consistent with the unrestricted HF study of 2D eliquid \cite{UHF} where larger $N$ has been accessed and the instability towards the Wigner crystal has been found at $r_s\sim 1.4$.

We prove that the Fermi gas ground state $|G\rangle$, assumed in the perturbative studies of the eliquid \cite{GB57,NP}, is unstable even in the limit $r_s\ll 1$. Consider a general HF state: $|G_U\rangle = \prod_\mathbf{s}\sum_\mathbf{p}\psi^+_\mathbf{p}U^+_{\mathbf{p}\mathbf{s}} |0\rangle $, created by the unitary rotation $U$ in the momentum space. It has the energy: $E_{HF}=(1/2m)\textrm{Tr}\left(\hat{\mathbf{p}}^2\rho\right)+$
\begin{equation}\label{HFEnergy}
+\frac{1}{2}\sum_\mathbf{q}U(\mathbf{q})\left[\textrm{Tr} \left(\delta_\mathbf{q}\rho\right) \textrm{Tr}\left(\delta_{-\mathbf{q}}\rho \right)- \textrm{Tr}\left(\delta_\mathbf{q} \rho\delta_{-\mathbf{q}}\rho \right) \right]
\end{equation}
where $\rho=UNU^+$ is the density matrix, $N$ is the diagonal occupation matrix: $N_\mathbf{s}=1$ and $N_\mathbf{p}=0$. The parameterization:
\begin{equation}
U=\left(\begin{array}{cc} \sqrt{1-z^+z} & z^+ \\ -z & \sqrt{1-zz^+} \end{array} \right),
\end{equation}
where $z$ is the rotation matrix from $\mathbf{s}$ to $\mathbf{p}$ space, allows one to expand the energy Eq.(\ref{HFEnergy}) in the vicinity of $|G\rangle$:
\begin{eqnarray}\label{HFEnergyExpand}
E_{HF}= \sum_{\mathbf{ps}}(\epsilon_{HF}(\mathbf{p})- \epsilon_{HF}(\mathbf{s})) z^+_{\mathbf{sp}} z_{\mathbf{ps}} + \nonumber\\ + \sum_{\mathbf{qss'}} (U(\mathbf{q})-U(\mathbf{s}-\mathbf{s}')) z^+_{\mathbf{s},\mathbf{s+q}} z_{\mathbf{s'+q},\mathbf{s'}} + \nonumber \\
\sum_{\mathbf{q}>0,\mathbf{ss'}} (U(\mathbf{q})-U(\mathbf{s}-\mathbf{s'}+\mathbf{q})) (z^+_{\mathbf{s},\mathbf{s+q}} z^+_{\mathbf{s'},\mathbf{s'-q}} + h.c. )
\end{eqnarray}
The instability momentum is $|\mathbf{Q}|=2p_F$ and the corresponding exciton wave function is $\psi(\mathbf{s})=z_{\mathbf{s+Q},\mathbf{s}}\sqrt{\mathcal{A}}$. For small deviations from the Fermi circle: $r=p_F-|\mathbf{s}|\ll p_F$, the wave function factorizes: $\psi(\mathbf{s})= \psi(r)\psi(\theta)$, where $\theta$ is the angle along the Fermi circle. The function $\psi(\theta)$ is positive within limits $\theta_{max}=R_0/p_F$ where $R_0=\sqrt{p_Fr_0}$ is the largest and $r_0$ is the smallest momentum scales of the function $\psi(r)$. Then, the energy Eq.(\ref{HFEnergyExpand}) becomes:
\begin{equation}\label{Functional}
\frac{E_{HF}}{2v_F}=\int_{r_0}^{R_0}\!\!\!\!\! dr\, r|\psi|^2-\frac{g}{2}\!\int\!\!\!\!\int_{r_0}^{R_0}\!\!\!\!\! drdr' \psi^*(r)\psi(r')\log\frac{R_0^2}{|r^2-r'^2|}
\end{equation}
where $g=e^2/(\pi v_F)\ll 1$ is the coupling. With the logarithmic accuracy the wave function depends on $\xi=\log(r/r_0)$. Provided $\sqrt{g}L\sim 1$, where $L=\log(R_0/r_0)$, the HF term in the kinetic energy is small $\sim g L$. Variation of the functional Eq.(\ref{Functional}) gives the equation for $\psi(r)$ with the solution: $\psi(r)=a\cos(\sqrt{g}\xi)/r$, for the eigenvalue $\sqrt{g}L=\pi/2$ and within the range $r_0\ll r\ll R_0$, where $a\sim\sqrt{r_0}$ is the normalization. The energy Eq.(\ref{Functional}) for this solution: $E_{HF}=2v_F a^2 \ \sin\left(2\sqrt{g} L\right)/\sqrt{g}$, becomes negative at $2\sqrt{g}L= \pi$ and is small $E_{HF}\sim -v_Fr_0$, for $r_s\ll 1$, with the instability momentum scale:
\begin{equation}\label{Scale}
r_0\approx p_F\exp\left(-\pi\sqrt{2\pi/r_s}\right)
\end{equation}
This proof can be extended to 3D and to eliquids with the spin. The concept of the 'normal' Fermi liquid does not apply to electron liquids, it is rather a quantum crystal.

In conclusion a new method for interacting electrons has been developed. It reveals that 2D electron liquid is a noncompensated quantum crystal at $r_s<10$ and is a fluid. Even the Wigner crystal, in the presence of pinning centers with the density $n_i$, flows. Indeed, a uniform force deforms the lattice and the stress builds up around the pinning cite making the quantum tunneling action $S(r_s)$ to be different for the lattice slip along or counter the force. The conductivity of the Wigner crystal is $\sigma_{C}\sim (e^2/\hbar) \sqrt{n/n_i}\exp(-S(r_s))$. The excitations of the eliquid falls into two types: the usual Landau quasiparticles in the vicinity of the Fermi surfaces and new Bose modes induced by the crystal order. The conductivity of the noncompensated quantum crystal includes the quasiparticle and the crystal terms: $\sigma=\sigma_F+\sigma_C$. The classical Hall effect of the noncompensated quantum crystal is $\sigma_H=enc/H$. Indeed, only the Lorentz and impurity forces are applied to electrons, but the latter is a viscous force with no component perpendicular to the current. The Hall effect refines into the negative part of the hole-like Fermi circles (involving the umklapp processes) and the crystal part from the whole Brillouin zone given by the Chern number. MIT in this picture is a disappearance of the Fermi circles as the crystal order parameter and the particle-hole gaps grow. In the MIT point the quantum crystal transforms into a regular Wigner crystal. For large crystal action $S(r_s)\sim\sqrt{r_s}$ the conductivity, $\sigma\approx \sigma_F$, will show the activated behaviour on the insulator side whereas on the metal side, where the charged carrier density $n(T)$ is low, it will be governed by the 2D classical plasma law at finite temperature: $\sigma_L\sim (e^2/\hbar) (e^2\sqrt{n(T)}/T)$. In the parallel magnetic field electrons in the crystal phase are easily polarized and all the tunneling probabilities, determining the quasiparticle mass and mobility, will decrease due to the Fermi sign repulsion.

I am grateful to G.M. Eliashberg and S.V. Iordanskii for their help, and to RFBR for support.


\begin{thebibliography}{22}
\bibitem{NP} P. Nozieres, 'Theory of Interacting Fermi Systems', (Benjamin, New York 1964)
\bibitem{landau56} L.D. Landau, Sov. Phys. JETP \textbf{3}, 920 (1957) 
\bibitem{QMC}  Y. Kwon, D. M. Ceperley, and R. M. Martin, Phys. Rev. B \textbf{53}, 7376 (1996); Phys. Rev. B \textbf{50}, 1684 (1994); Phys. Rev. B \textbf{48}, 12037 (1993)
\bibitem{QMC1} C. Attaccalite, S. Moroni, P. Gori-Giorgi and G. B. Bachelet, Phys. Rev. Lett. \textbf{88}, 256601 (2002)
\bibitem{krav94} S. V. Kravchenko, G. V. Kravchenko, J. E. Furneaux, V. M. Pudalov, and M. D'Iorio, Phys. Rev. B \textbf{50}, 8039 (1994); E. Abrahams, S. V. Kravchenko and M. P. Sarachik, Rev. Mod. Phys. \textbf{73}, 251 (2001)
\bibitem{AGD} A. A. Abrikosov, L. P. Gor'kov and I. E. Dzyaloshinski, 'Methods of Quantum Field Theory in statistical Mechanics' (Dover, New York 1963)
\bibitem{pomer} I.Ya. Pomeranchuk, Sov. Phys. JETP \textbf{8}, 361 (1959) 
\bibitem{AL69} A. F. Andreev and I. M. Lifshitz, Sov. Phys. JETP \textbf{29}, 1107 (1969)
\bibitem{Yang} A.P. Young, in 'Strongly interacting fermions and high Tc superconductivity', eds. B B. Doucot and J. Zinn-Juistin (Elsevier 1995)
\bibitem{UHF} J.R. Trail, M.D. Towler and R.J. Needs, Phys. Rev. B \textbf{68}, 045107 (2003)
\bibitem{GB57} M. Gell-Mann and K. A. Brueckner, Phys. Rev. \textbf{106}, 364 (1957)
\end{thebibliography}
\end{document}